\documentclass[11pt,preprintnumbers,titlepage,nofootinbib]{revtex4-2}
\usepackage[latin9]{inputenc}
\usepackage{amsmath}
\usepackage{amssymb}
\usepackage[unicode=true,
 bookmarks=false,
 breaklinks=false,pdfborder={0 0 1},backref=section,colorlinks=false]
 {hyperref}
\hypersetup{
 colorlinks,linkcolor=blue,citecolor=blue,urlcolor=blue}

\makeatletter

\usepackage{mathrsfs}
\usepackage{graphicx}
\usepackage{wasysym}
\usepackage{array}
\usepackage{multirow}
\usepackage{wasysym}
\usepackage{wasysym}
\usepackage{amsfonts}
\usepackage{subfigure}
\usepackage{cleveref}
\usepackage{listings}
\usepackage{xcolor}
\usepackage{array}
\usepackage{url}
\usepackage{color}
\usepackage{subfigure}

\@ifundefined{textcolor}{}{
\definecolor{BLACK}{gray}{0}
\definecolor{WHITE}{gray}{1}
\definecolor{RED}{rgb}{1,0,0}
\definecolor{GREEN}{rgb}{0,1,0}
\definecolor{BLUE}{rgb}{0,0,1}
\definecolor{CYAN}{cmyk}{1,0,0,0}
\definecolor{MAGENTA}{cmyk}{0,1,0,0}
\definecolor{YELLOW}{cmyk}{0,0,1,0}
\definecolor{ballblue}{rgb}{0.13, 0.67, 0.8}
\definecolor{bleudefrance}{rgb}{0.19, 0.55, 0.91}
\definecolor{blue(ncs)}{rgb}{0.0, 0.53, 0.74}
\definecolor{darkpastelgreen}{rgb}{0.01, 0.75, 0.24}
\definecolor{darkspringgreen}{rgb}{0.09, 0.45, 0.27}
\definecolor{denim}{rgb}{0.08, 0.38, 0.74}
\definecolor{electricviolet}{rgb}{0.56, 0.0, 1.0}
}

\makeatother

\begin{document}
\preprint{CTP-SCU/2025010}
\title{Observational Signatures of Janis-Newman-Winicour Strongly Naked Singularity}



\author{Zeqian Zhang}
\email{zhangzeqian@stu.scu.edu.cn}

\affiliation{Center for Theoretical Physics, College of Physics, Sichuan University,
Chengdu, 610064, China}
\begin{abstract}
This paper explores the unique observational signatures of accretion onto a Janis-Newman-Winicour (JNW) strongly naked singularity, particularly in the absence of a photon sphere. The surrounding spacetime of such a singularity exhibits pronounced reflective properties, causing light rays traveling in its vicinity to undergo reflection and produce paired imaging trajectories in both accretion disk and hot spot models. Our simulations reveal that this reflection effect generates additional images in the gravitational lensing patterns and significantly alters the temporal brightness profiles of the lensed images. These reflection-induced paired images, together with their inclination-dependent behavior, offer distinct observational signatures that could potentially distinguish a JNW strongly naked singularity from a black hole based on future high-resolution observations.
\end{abstract}
\maketitle
\tableofcontents{}
\section{Introduction}

\label{sec:Introduction}

The study of compact astrophysical objects such as black holes and naked singularities represents a critical frontier in our understanding of general relativity and gravitational physics. Black holes, long considered the inevitable outcome of gravitational collapse, are characterised by the presence of an event horizon-a boundary beyond which no information can escape. In contrast, naked singularities are hypothetical solutions to Einstein's field equations where the central singularity remains exposed to distant observers, without the shielding of an event horizon. Owing to the existence of photon spheres in many naked singularity models, these objects can reproduce observational features traditionally attributed to black holes. This possibility motivates detailed investigations into their gravitational lensing properties and distinctive observable imprints \cite{Virbhadra:2002ju,Virbhadra:2007kw,Gyulchev:2008ff,Sahu:2012er,Shaikh:2019itn,Paul:2020ufc,Wyman:1981bd,Tsukamoto:2021fsz,Wang:2023jop,Chen:2023trn,Schmidt:2008hc,Guzik:2009cm,Liao:2015uzb,Goulart:2017iko,Nascimento:2020ime,Islam:2021ful,Tsukamoto:2021caq,Junior:2021svb,Olmo:2021piq,Ghosh:2022mka}.
\par
The ability to observationally distinguish a naked singularity from a black hole carries profound implications, extending beyond theoretical considerations. The recent Event Horizon Telescope (EHT) observations of black hole shadows have provided unprecedented insights into the near-horizon regions of compact objects \cite{Akiyama:2019cqa,Akiyama:2019brx,Akiyama:2019sww,Akiyama:2019bqs,Akiyama:2019fyp,Akiyama:2019eap,Akiyama:2021qum,Akiyama:2021tfw,EventHorizonTelescope:2022xnr,EventHorizonTelescope:2022vjs,EventHorizonTelescope:2022wok,EventHorizonTelescope:2022exc,EventHorizonTelescope:2022urf,EventHorizonTelescope:2022xqj}, creating new opportunities to confront theoretical models with empirical data. However, the possibility remains that certain horizonless ultra-compact objects, such as naked singularities, might mimic black hole shadows via the formation of photon spheres, complicating the task of distinguishing them observationally \cite{Narayan:2019imo,Shaikh:2018lcc,Zeng:2020dco,Zeng:2020vsj,Qin:2020xzu,Luminet:1979nyg,Beckwith:2004ae,Gralla:2019xty,Dokuchaev:2019pcx,Peng:2020wun,He:2021htq,Eichhorn:2021iwq,Li:2021riw,Gan:2021pwu,Gan:2021xdl}.
\par
Among the proposed naked singularity solutions, the Janis-Newman-Winicour (JNW) spacetime, involving a minimally coupled massless scalar field, has attracted considerable attention \cite{Chowdhury:2011aa,Janis:1968zz,Joshi:2019aa}. This solution challenges conventional wisdom by admitting the formation of naked singularities under specific conditions, offering a valuable theoretical setting to explore possible violations of the cosmic censorship conjecture. In the absence of a photon sphere, the JNW strongly naked singularity exhibits distinctive gravitational lensing properties. Photons propagating in its vicinity may undergo reflection due to the divergent effective potential near the singularity, generating paired imaging trajectories and, in some cases, additional images. These reflection-induced features directly affect the morphology and brightness distribution of observed images, notably those of accretion disks. The prospect of such multi-image structures, including inversion phenomena arising from light path reflection, presents a promising avenue for identifying observational signatures that differentiate a JNW naked singularity from a black hole.
\par
Previous studies have investigated gravitational lensing by naked singularities in both Janis-Newman-Winicour and Born-Infeld spacetimes, primarily focusing on general lensing properties and photon trajectories, including the possibility of light traversing regularised singularities \cite{Patil:2011aa,Gyulchev:2019tvk,Sau:2020xau,Gyulchev2020,Chauvineau:2022bzg,Chen:2023uuy,Chen:2023trn}. Building on this foundation, the present work concentrates on the specific observational imprints of a strongly naked JNW singularity. Particular attention is devoted to the reflective character of the spacetime, and how this affects the structure of images formed by accretion disks and dynamic hot spots orbiting near the singularity.
\par
A particularly promising approach to probing the strong gravity regime around compact objects involves the study of localised, orbiting hot spots. General Relativistic Magnetohydrodynamical (GRMHD) simulations and semi-analytic models suggest that magnetic reconnection events and flux eruptions within magnetised accretion disks can generate transient hot spots orbiting near the innermost stable circular orbit (ISCO) of supermassive black holes \cite{Hamaus:2008yw, Trippe2007, Broderick2006}. Observationally, evidence for such orbiting hot spots has been repeatedly reported in the vicinity of Sgr A* \cite{Witzel:2020yrp,Michail:2021pgd,GRAVITY:2021hxs}, and more recently, a candidate hot spot signature has been identified in unresolved light curve data obtained by the Event Horizon Telescope \cite{EventHorizonTelescope:2022exc}. Due to their compact size, proximity to the central object, and distinctive brightness variability, these hot spots offer an effective means of testing the spacetime geometry in the immediate vicinity of compact objects \cite{Wielgus:2022heh,abuter2018detection}.
\par
The subsequent sections of this paper are structured as follows: In Section \ref{sec:Setup}, we briefly review the Janis-Newman-Winicour (JNW) strongly naked singularity spacetime and discuss the properties of light propagation in the absence of a photon sphere. Section \ref{sec:Observational signatures of JNW strongly naked singularity} is dedicated to the numerical simulation framework, including ray-tracing methods applied to three complementary configurations: \ref{sec:Celestial sphere model} a celestial sphere background,  \ref{sec:Accretion disk model} a thin accretion disk, and \ref{sec:Hot spot model} a hot spot model. This section also presents the resulting time-integrated images, temporal brightness variations, and centroid trajectories, with emphasis on identifying reflection-induced image pairs, inversion effects, and their dependence on the observer's inclination. Finally, Section \ref{sec:CONCLUSIONS} summarises our main conclusions and discusses the observational implications of these findings. We adopt natural units with $G = c = 1$ throughout this paper \cite{Abbott:2016blz}.

\section{Setup}

\label{sec:Setup}

In this section, we provide a concise insight into the JNW metric, as well as the properties of the corresponding spacetime. Subsequently, our focus shifts to discussing stable circular orbits around the naked singularity for the purpose of facilitating the follow-up study.

\subsection{JNW Metric}

One of the simplest solutions to Einstein's equation is the static spherically symmetric vacuum solution, namely the Schwarzschild solution. Now we consider a massless scalar field which is minimally coupled to gravity, whose action is given by
\begin{equation}
S=\int\mathrm{d}^4x\sqrt{-g}\left(\frac{1}{16\pi G}R-\frac{1}{2}g^{\mu\nu}\partial_{\mu}\phi\partial_{\nu}\phi\right).
\end{equation}
By solving the corresponding Einstein's equation
\begin{equation}
R_{\mu\nu}-\frac{1}{2}Rg_{\mu\nu}=8\pi GT_{\mu\nu},\quad T_{\mu\nu}=\partial_{\mu}\phi\partial_{\nu}\phi-\frac{1}{2}g_{\mu\nu}g^{\rho\sigma}\partial_{\rho}\phi\partial_{\sigma}\phi,
\end{equation}
we obtain a static spherically symmetric solution
\begin{equation}
\mathrm ds^2=-\left(1-\frac{r_c}{r}\right)^{\gamma}\mathrm dt^2+\left(1-\frac{r_c}{r}\right)^{-\gamma}\mathrm dr^2+\left(1-\frac{r_c}{r}\right)^{1-\gamma}r^2\left(\mathrm d\theta^2+\text{sin}^2\theta \mathrm d\phi^2\right),
\end{equation}
known as the Janis-Newman-Winicour metric. The scalar field $\phi$ is given by
\begin{equation}
\phi\left(r\right)=\frac{q}{r_c}\text{ln}\left(1-\frac{r_c}{r}\right).
\end{equation}
And $r_c$ is related to the scalar charge $q$ and the ADM mass $M$, given by
\begin{equation}
r_c=2\sqrt{M^2+q^2}.
\end{equation}
Thus we have $0\leq\gamma\leq1$ and $r_c\gamma=2M$, from which we can easily obtain that a smaller $\gamma$ corresponds to a larger magnitude of the scalar field. And when $\gamma=1$, the metric degenerates to the Schwarzschild metric. Besides, there is a naked curvature singularity at $r=r_c$, and therefore we confine the observer within the region of $r>r_c$.
\par
For a particle with four-velocity vector $U^{\alpha}=\frac{dx^{\alpha}}{d\lambda}$, where $\lambda$ is the affine parameter along the geodesic. Since the JNW metric is spherically symmetric, we consider geodesics on the equatorial plane, which means that $\theta=\frac{\pi}{2}$ and $U^{\theta}=0$. Hence, we have two conserved quantities to characterise the geodesic equations, namely
\begin{equation}
\begin{aligned}
E&=-g_{tt}U^t=-p_t,\\
L&=g_{\phi\phi}U^{\phi}=p_{\phi},
\end{aligned}
\end{equation}
which represent energy and angular momentum respectively. By virtue of the normalization $U^{\alpha}U_{\alpha}=-1$, we obtain the relationship between the effective potential $V_{eff}$ and energy,
\begin{equation}
\dot{r}^2+V_{eff}=E^2.
\end{equation}
\par
For a photon travelling along a null geodesic, its effective potential is given by
\begin{equation}
V_{eff}=\frac{L^2}{r^2\left(1-\frac{r_c}{r}\right)^{1-2\gamma}}.
\end{equation}
In the circumstances of $0.5<\gamma\leq1$, the effective potential encounters a local maximum at which the photon sphere is located. Solving $V_{eff}^{\prime}=0$ and $V_{eff}^{\prime\prime}<0$, we obtain the radius of the photon sphere,
\begin{equation}
r_{ph}=r_c\left(\gamma+\frac{1}{2}\right).
\end{equation}
Since $r_{ph}>r_{c}$ in these cases, the naked singularity is covered by a photon sphere. Therefore, we refer to this type of naked singularity as a weakly naked singularity.
\par
On the other hand, when $0\leq\gamma\leq0.5$, no maximum occurs within the region of $r>r_c$ and the effective potential tends to infinity at the singularity. As a result, no photon sphere exists in the spacetime. Hence, we refer to this type of naked singularity as a strongly naked singularity. Since a strongly naked singularity may lead to significant differences in comparison with a weakly naked one, the following part will focus on the JNW metric with  $0\leq\gamma\leq0.5$.

\subsection{Stable circular orbits in JNW spacetime}

For massive particles travelling around the naked singularity, we can expect that the parameter $\gamma$ shall have an impact on their behaviour. Now consider a timelike geodesic on the equatorial plane, the effective potential $V_{eff}$ can be described as
\begin{equation}
V_{eff}=\left(1-\frac{r_c}{r}\right)^{\gamma}\left(1+\frac{L^2}{r^2\left(1-\frac{r_c}{r}\right)^{1-\gamma}}\right).
\end{equation}
Since the focus is on the circular geodesics around the naked singularity, it is necessary to put some constraints on the equation of motion, which is given by
\begin{equation}
U^r=\dot{U}^r=0.
\end{equation}
where the dotted quantity represents the derivative concerning the affine parameter, namely its proper time $\tau$. Thus we have $V_{eff}=E^2$ and $\frac{\partial V}{\partial r}=0$.
Imposing the constraints on the effective potential, we obtain the expressions for the conserved energy and angular momentum (detailed information in \cite{Chowdhury:2011aa})
\begin{equation}
\begin{aligned}
E^2&=\left(1-\frac{r_c}{r}\right)^{\gamma}\frac{2r-r_c(\gamma+1)}{2r-r_c(2\gamma+1)},\\
L^2&=\left(1-\frac{r_c}{r}\right)^{1-\gamma}\frac{\gamma r_c r^2}{2r-r_c(2\gamma+1)}.
\end{aligned}
\end{equation}
The complete expression for effective potential will be obtained by substituting $L^2$ into $V_{eff}$.
\par
Due to the fact that we are looking for stable circular orbits, it is a must that the effective potential admits a minimum. Therefore, we have
\begin{equation}
\frac{\partial^2}{\partial r^2}V_{eff}>0.
\end{equation}
\begin{figure}[htbp]
\centering
\includegraphics[scale=0.7]{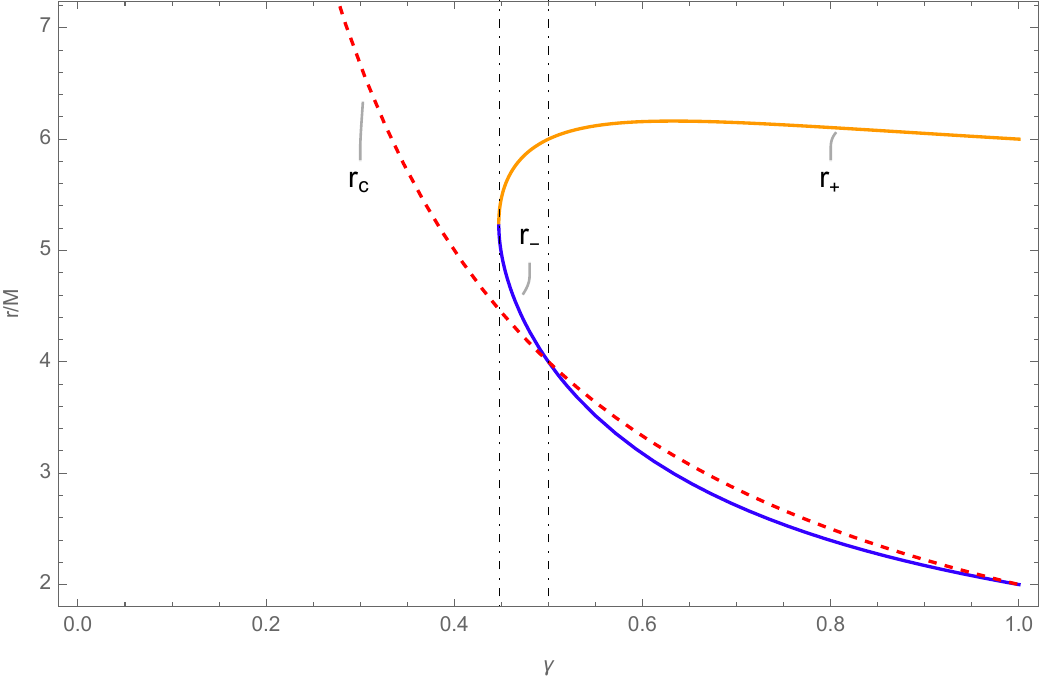}
\caption{The radius of MSCOs $r_{\pm}$ (orange line and blue line, respectively) and the naked singularity $r_c$ (dashed red line) are shown as the parameter $\gamma$ varies. The structure of circular geodesics around the naked singularity will be different in three intervals, namely $\left(0,\frac{1}{\sqrt{5}}\right)$, $\left(\frac{1}{\sqrt{5}},0.5\right)$ and $\left(0.5,1\right)$.}
\label{Figure 1}
\end{figure}
Solving the inequality, we obtain the conditions for stable circular orbits,
\begin{equation}
r>r_+\quad\text{or}\quad r<r_-,
\end{equation}
where $r_{\pm}$ is the marginally stable circular orbits (MSCOs), given in \cite{Chowdhury:2011aa, Joshi:2019aa} by
\begin{equation}
r_{\pm}=\frac{r_c}{2}\left(1+3\gamma\pm\sqrt{5\gamma^2-1}\right).
\end{equation}
Additionally, the orbits should be confined to the region of $r>r_c$.
\par
In FIG. \ref{Figure 1} the radius of MSCOs $r_{\pm}$ and the naked singularity $r_c$ are shown as the parameter $\gamma$ varies. Since the constraint $r>r_c$ is imposed, the inner MSCO, namely $r_-$, will be covered by the naked singularity when $\gamma>0.5$. Therefore, there exists an innermost stable circular orbit (ISCO) outside the naked singularity, in which circumstance, the radius of the ISCO is given by $r_{ISCO}=r_+$. Nevertheless, as for the strongly naked singularity with $0 \leq \gamma < 0.5$, when $\gamma$ takes values between $1/\sqrt{5}$ and $0.5$, we have $r_c < r_- < r_+$. In this case, stable circular orbits outside the naked singularity exist only in the regions where $r \leq r_-$ or $r \geq r_+$. When $\gamma$ decreases below $1/\sqrt{5}$, stable circular orbits are permitted at any radius in the spacetime.

\section{Observational signatures of JNW strongly naked singularity}

\label{sec:Observational signatures of JNW strongly naked singularity}

Equipped with the expression for the JNW metric and stable circular orbits, we can visualise observational signatures of the JNW strongly naked singularity utilising ray tracing and numerical simulation. In this section, three different models are applied, namely the celestial sphere model, accretion model and hot spot model, which enable us to gain deeper insight into the gravitational lensing by a strongly naked singularity.

\subsection{Celestial sphere model}

\label{sec:Celestial sphere model}

To illustrate the light propagation in the general spacetime outside the JNW naked singularity, a celestial sphere functioning as a illuminant is set at $r=25M$ with a observer located at $x^{\mu}_{obs}=(t, r, \theta, \phi)=(0, 10M, \pi/2, \pi)$. The celestial sphere is divided into four quadrants, which are differently colored, while a white dot is placed in front of the observer to demonstrate the structure of the Einstein ring. In addition, we highlight a grid of equidistantly separated black lines, with an interval of $\pi/18$, to represent constant longitude and latitude. To generate observational images, we vary the observer's viewing angle and numerically integrate $2000\times2000$ photon trajectories until they intersect with the celestial sphere. One can refer to \cite{Chen:2023trn} for a more detailed illustration.
\begin{figure}[htbp]
\centering
\subfigure
{
\begin{minipage}[b]{.3\linewidth}
\centering
\includegraphics[scale=0.125]{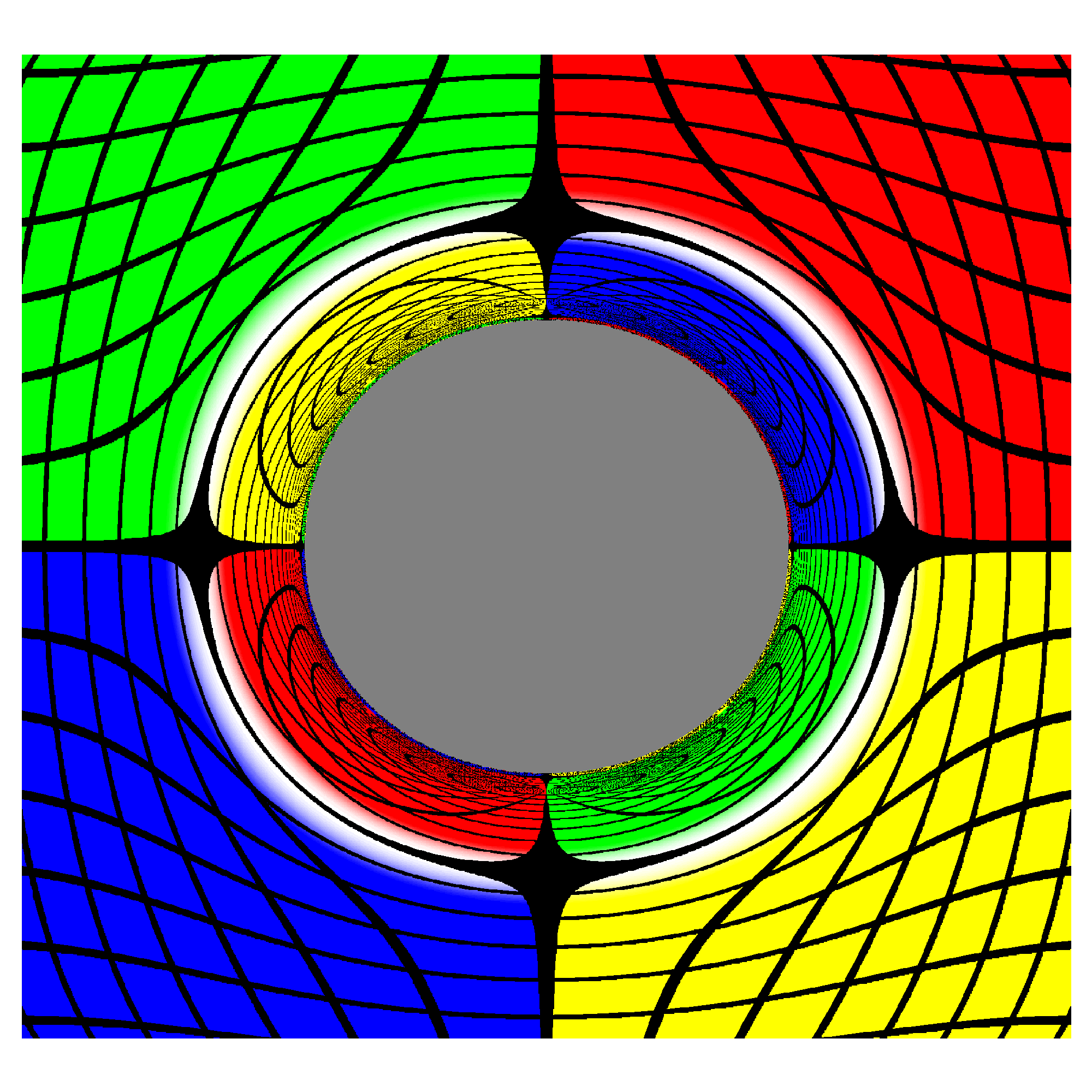}
\end{minipage}
}
\subfigure
{
\begin{minipage}[b]{.3\linewidth}
\centering
\includegraphics[scale=0.125]{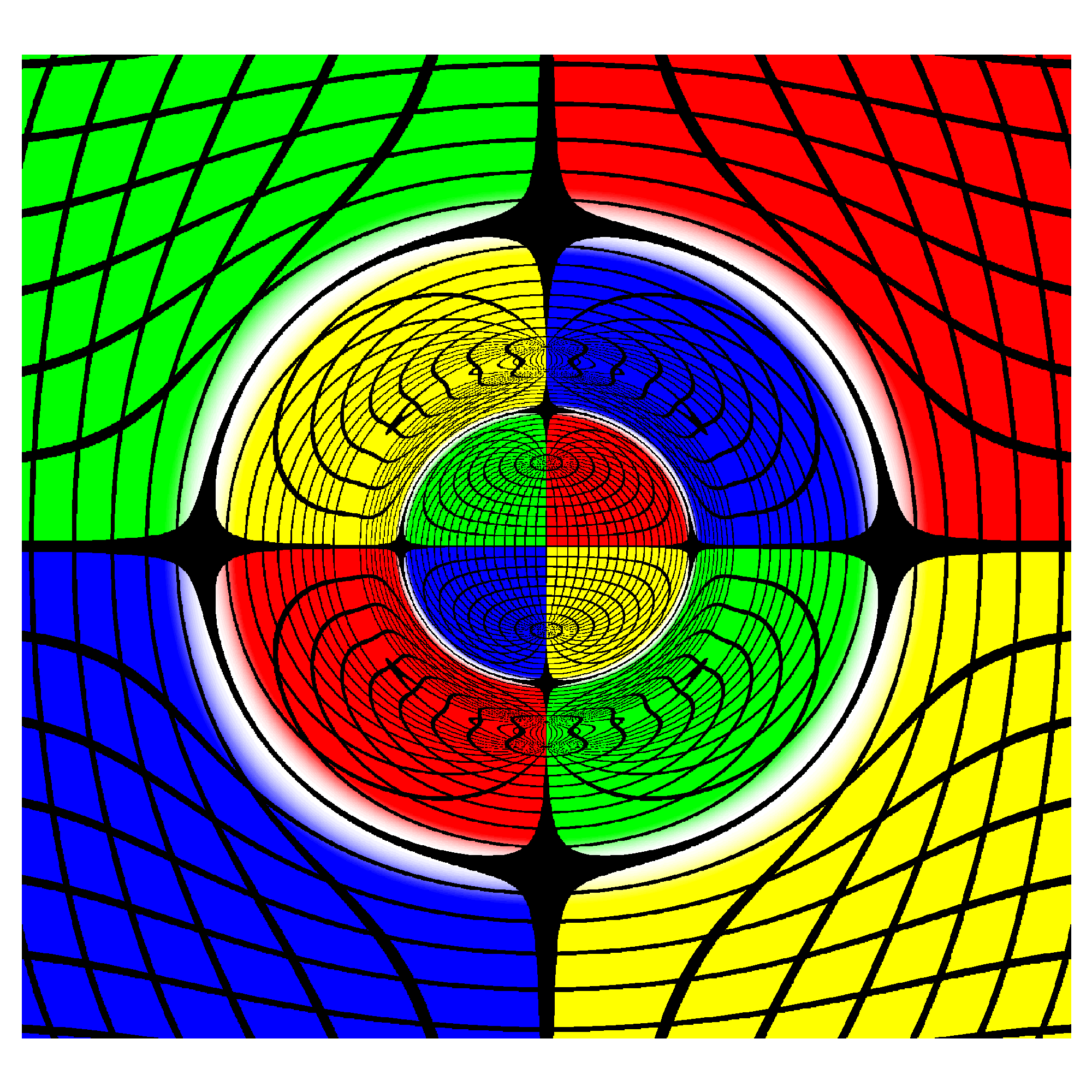}
\end{minipage}
}
\subfigure
{
\begin{minipage}[b]{.3\linewidth}
\centering
\includegraphics[scale=0.125]{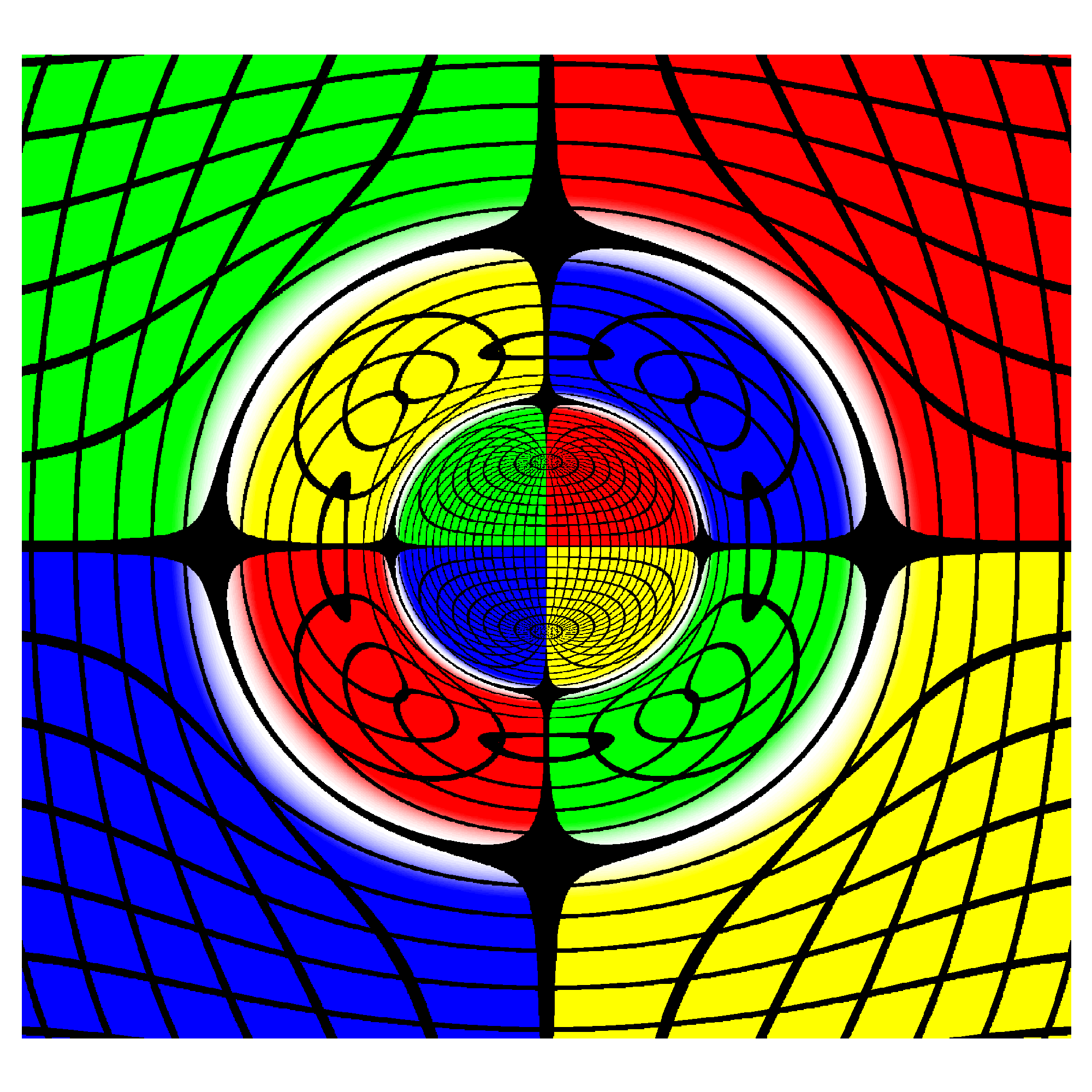}
\end{minipage}
}
\caption{Images of a celestial sphere located at $r=25M$ in the JNW metric with the parameter $\gamma=1$ (\textbf{Left}), 0.45 (\textbf{Middle}) and 0.4 (\textbf{Right}). In the cases of $\gamma=1$, the graph displays the celestial sphere of the Schwarzschild black hole, while the other two display those of the JNW strongly naked singularity. The observer is located at $x^{\mu}_{obs}=(t, r, \theta, \phi)=(0, 10M, \pi/2, \pi)$, and the field of view spans $2\pi/3$. Only one Einstein ring occurs in the left panel, while the middle and right panels exhibit two Einstein rings. In the $\gamma=1$, namely Schwarzschild black hole case, the image draws up a grey area as the black hole shadow, which is a dark patch perceived by the observer. From the middle and right panels, which exhibit similar characteristics compared with the Schwarzschild black hole, one can note that there exists an additional image in the graph of the celestial sphere at a certain radius, which depends on the value of $\gamma$.}
\label{Figure 2}
\end{figure}
\par
FIG. \ref{Figure 2} presents the celestial sphere images in the spacetimes of a Schwarzschild black hole and JNW strongly naked singularities. In the left panel, corresponding to $\gamma=1$, the image shows a black hole shadow in grey, enclosed by an Einstein ring. Several high-order gravitational lensing images appear near the photon sphere at the shadow's edge. In the middle panel, for a JNW naked singularity with $\gamma=0.45$, the photon sphere and shadow disappear. Instead, a high-order-like image emerges at the centre, accompanied by two Einstein rings. Within the region between these rings, additional images appear - a feature absent in the Schwarzschild case. The right panel shows a similar pattern. Two Einstein rings are present, and extra images emerge in the intermediate region, with their positions depending on the value of $\gamma$.
\par
\begin{figure}[htbp]
\centering
\includegraphics[scale=0.9]{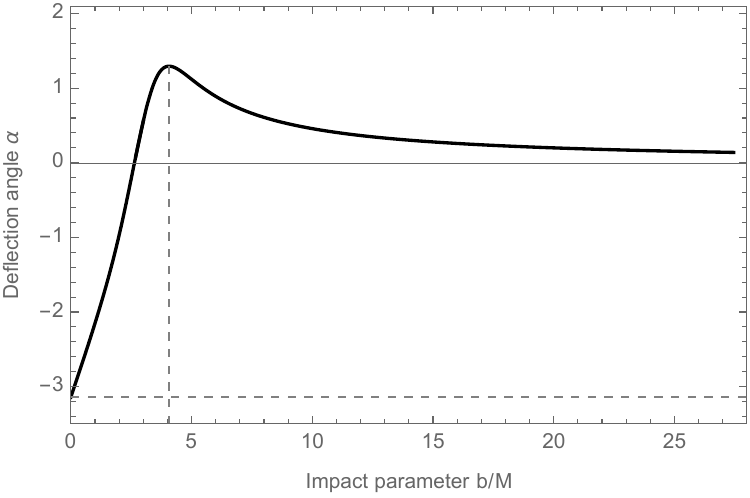}
\caption{The deflection angle $\alpha(b)$ as a function of the impact parameter $b$ for a JNW naked singularity with $\gamma=0.4$. The deflection angle attains its maximum value $\alpha_{\mathrm{max}} = \alpha(b_m)$ at $b = b_m$, while approaching $\alpha(0) = -\pi$ as $b \to 0$.}
\label{Figure 3}
\end{figure}

To figure out the high-order-like images and the emergence of additional images, we seek clarification towards the deflection angle, which is a function of the impact parameter $b$. The deflection angle is given by
\begin{equation}
\alpha(b)=I(b)-\pi,
\end{equation}
where $I(b)$ is defined in \cite{Chen:2023trn, Chen:2023uuy, Shaikh:2019itn}, which is numerically integrated. As the impact parameter varies from $0$ to infinity, the deflection angle $\alpha(b)$ reaches its maximum $\alpha_{max}=\alpha(b_m)$, shown in FIG. \ref{Figure 3}. Additionally, its value at $b=0$ and infinity are, respectively, $\alpha(0)=-\pi$ and $\alpha(\infty)=0$. Therefore, it can be concluded that the high-order-like images originate from light rays undergoing reflection in the vicinity of the naked singularity. These additional images correspond to regions of the celestial sphere located behind the observer. The discontinuity in the image distribution marks the critical point where the reflection takes place, resulting in a reversal of the light ray's propagation direction relative to the observer's line of sight, thereby producing an inverted image pattern. Comparing the middle and right panels of FIG. \ref{Figure 2}, it is evident that the radial position of these additional images depends on the value of $\gamma$. As $\gamma$ decreases, the corresponding $b_m$ increases while the maximum deflection angle $\alpha(b_m)$ decreases. Consequently, for sufficiently small $\gamma$, no Einstein ring will form on the celestial sphere of the JNW naked singularity.

\subsection{Accretion disk model}

\label{sec:Accretion disk model}

Having clarified the celestial sphere in JNW spacetime, in this section we will focus on the property of the imaging of the accretion disk around the JNW strongly naked singularity. We establish the same setup as that in \cite{Sau:2020xau} for the accretion disk, which is considered to be geometrically thin and optically thick. The accretion is generated by a collection of luminous particles orbiting the naked singularity. To maintain the stability of the accretion disk, the orbiting particles should be constrained on the stable circular orbits mentioned in section \ref{sec:Setup}. Assuming the accretion process exists on all circular orbits in JNW spacetime, the distinct structures of the accretion disk are portrayed in \cite{Gyulchev2020}, which can extend to infinity. We set the accretion disk on the equatorial plane as well as an observer at the north pole of the naked singularity with $r_{obs}= 20 M$ (studies of accretion disks with tilted orientations relative to the observer have been extensively discussed in \cite{Gyulchev2020, Deliyski:2024aa}). To generate provoking observational images, we vary the observer's viewing angle and numerically integrate $1000\times1000$ photon trajectories. Upon the trajectory intersecting with the accretion disk, the corresponding viewing angle will be assigned with a flux, whose value is a function of $\gamma$ and the radius $r$ (detailed in \cite{Chen:2023qic}).
\begin{figure}[htbp]
\centering
\subfigure
{
\begin{minipage}[b]{.3\linewidth}
\centering
\includegraphics[scale=0.4]{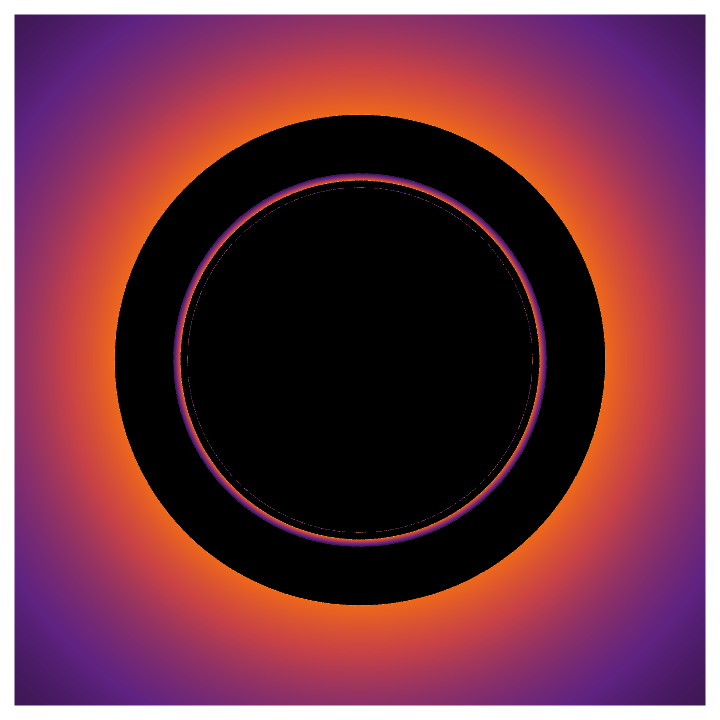}
\end{minipage}
}
\subfigure
{
\begin{minipage}[b]{.3\linewidth}
\centering
\includegraphics[scale=0.4]{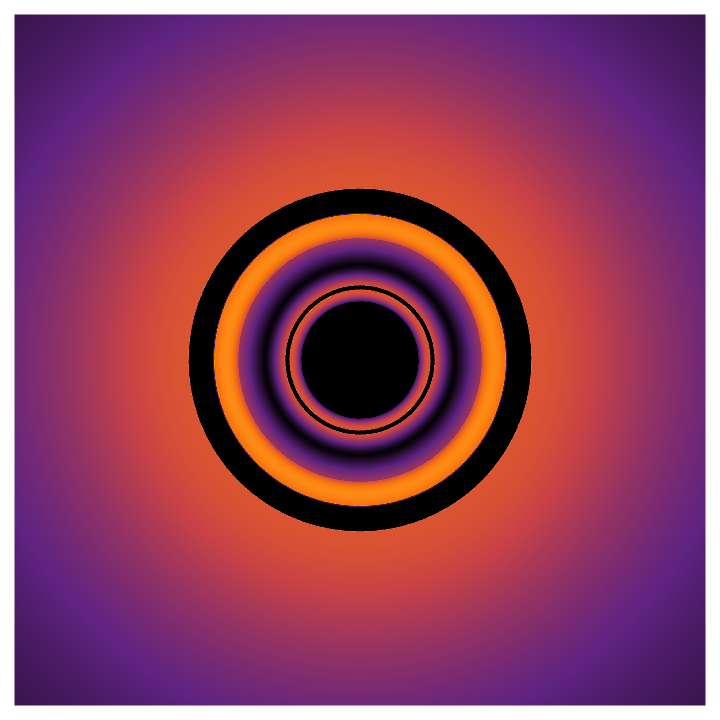}
\end{minipage}
}
\subfigure
{
\begin{minipage}[b]{.3\linewidth}
\centering
\includegraphics[scale=0.4]{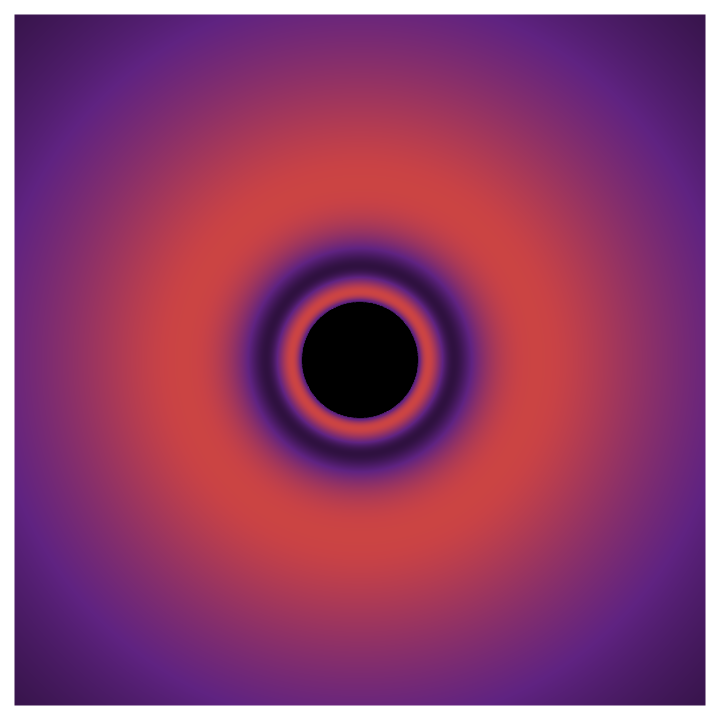}
\end{minipage}
}
\caption{Images of an accretion disk fill all stable circular orbits in JNW spacetime with $\gamma=1$ (\textbf{Left}), $0.45$ (\textbf{Middle}), and $0.4$ (\textbf{Right}). The observer is located at the north pole of the JNW naked singularity with $r_{obs}= 20 M$, and the field of view spans $1$ radian. In the $\gamma=1$, namely Schwarzschild black hole case, the image depicts the primary and secondary images of the accretion disk, as well as a photon ring. In the middle panel, the accretion disk viewed by the observer manifests four ring-shaped images. Within the region between the second and third rings counted from the outside in, there exists a dim image, while the other two intervals are devoid of any visible images. The right panel with $\gamma=0.4$ exhibits an image with an interval where the light rays are much fainter as well. Nevertheless, unlike the scenario with $\gamma=0.45$, no faults are evident in this image.}
\label{Figure 4}
\end{figure}
\par
FIG. \ref{Figure 4} depicts the observational images of an accretion disk of a Schwarzschild black hole and a JNW strongly naked singularity. The left panel reveals the case of a Schwarzschild black hole ($\gamma=1$), where the accretion disk extends from $r_{ISCO}=6M$ to infinity. It is composed of a large primary image and a thin ring-shaped image, namely the secondary image, which is generated by light bending around the black hole and ultimately intersecting with the accretion disk. One can also perceive that inside the secondary image, there exists a faintly visible photon ring, which is consistent with the previous discussion.
\par
The middle panel displays the accretion disk of a JNW strongly naked singularity with $\gamma=0.45$, in which case, the accretion disk exists from the singularity to infinity, except for an interval between the two MSCOs. Unlike the former one, the image is much more complex, which is composed of four ring-shaped images. The largest and outermost ring is the primary image, while the left three ring-shaped images were placed irregularly. Intriguingly, within the region between the second and third rings counted from the outside in, a dim image emerges. Yet, no images are visible in the other two intervals. Thus, we can speculate that the dim image is generated by light rays which undergo the process of reflection. More detailed clarification will be elaborated in the subsequent context.
\par
The right panel, corresponding to $\gamma=0.4$, reveals the accretion disk which fills the equatorial plane. Similar to but simpler than the circumstance of $0.45$, the image includes a large primary and a dim image as well, inside of which, there exists a small ring-shaped image. As we have discussed previously, since the light rays can be reflected in the vicinity of the naked singularity, the inner ring is the image of the relatively outer part of the accretion disk. Hence, the faintly visible image arises from reflected light rays sweeping across the accretion disk. As these rays approach the critical reflection region near the singularity, their apparent angular velocity in the observer's sky increases rapidly, causing the corresponding image to become increasingly compressed and dimmer until it eventually fades from view.
\begin{figure}[htbp]
\centering
\subfigure
{
\begin{minipage}[b]{.45\linewidth}
\centering
\includegraphics[scale=0.55]{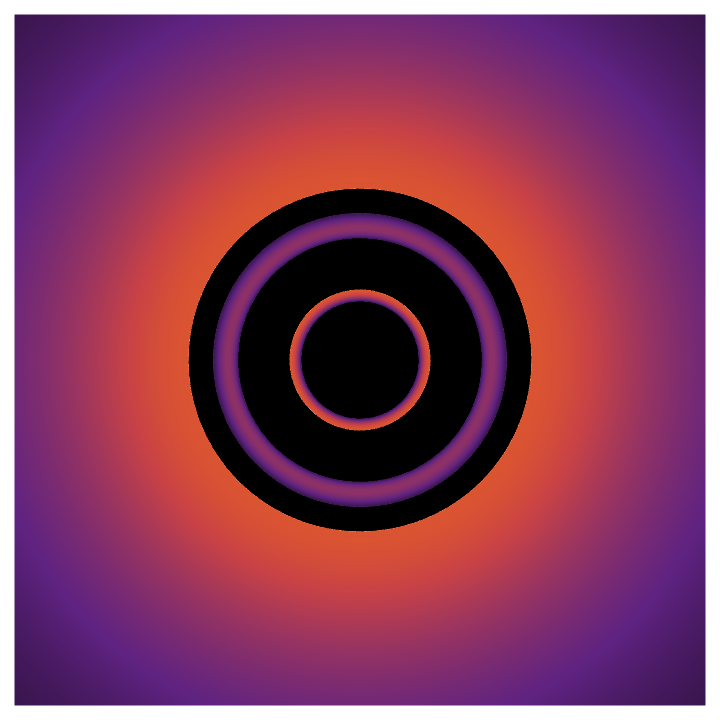}
\end{minipage}
}
\subfigure
{
\begin{minipage}[b]{.45\linewidth}
\centering
\includegraphics[scale=0.55]{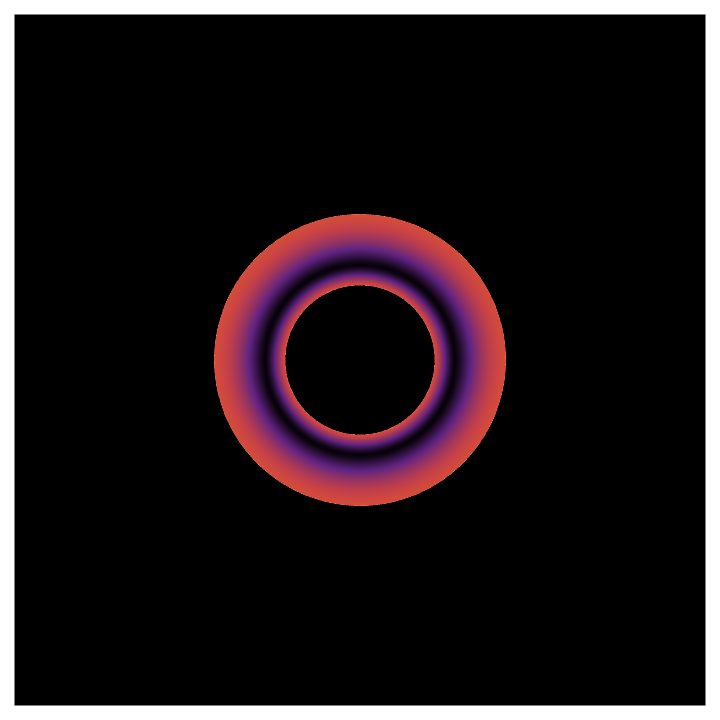}
\end{minipage}
}
\caption{Separated images of the outer (\textbf{Left}) and inner (\textbf{Right}) accretion disk in the circumstance of $\gamma=0.45$. We employ the same observer setup as depicted in FIG. \ref{Figure 4} with a field of view of $1$ radian. In the left panel, the outer accretion disk manifests three ring-shaped images, the second of which is fainter. Remarkably, no images are portrayed within the two intervals. The right panel exhibits the inner disk viewed by the observer, revealing only two ring-shaped images. In this case, the dim image emerges between the two rings. }
\label{Figure 5}
\end{figure}
\par
Building upon the preceding discussion, particularly highlighting the complexity of the case of $\gamma=0.45$, which was briefly expounded upon with the promise of further elaboration. Now, we delve deeper into this complexity by introducing separated images of the outer and inner accretion disk with $\gamma=0.45$ in FIG. \ref{Figure 5}, providing a comprehensive understanding of its intricacies.
\par
The left panel of FIG. \ref{Figure 5} exhibits the image of the outer accretion disk, which extends from $r_+$ to infinity. Unlike the middle panel of FIG. \ref{Figure 4}, only three ring-shaped images are portrayed in this case. Hence, we can infer that the three rings counted from the outside in are, respectively, the primary and secondary image of the outer accretion, and the image formed due to the reflected light rays. On the other hand, the right panel manifests the image of the inner accretion disk, which only exists within the interval between the naked singularity and $r_-$. With two ring-shaped images and a dim image emerging between the two rings, its genesis can be elaborated by employing the same clarification as that of the right panel of FIG. \ref{Figure 4}, namely the $\gamma=0.4$ case, just replacing for inner accretion disk. Overlaying the two separated images, we obtain the middle panel of FIG. \ref{Figure 4}. Thus, the two intervals devoid of any visible images correspond to the region between the inner and outer accretion disk.

\subsection{Hot spot model}

\label{sec:Hot spot model}

Observational data indicate that flares frequently occur on the accretion disks of compact objects \cite{Hamaus:2008yw}. These flares are widely interpreted as arising from localised over-dense regions, or hot spots, orbiting close to the innermost stable circular orbit, where synchrotron radiation is emitted by energetic particles \cite{Hamaus:2008yw, Trippe2007, Broderick2006}. Motivated by this, we investigate in this section the observational signatures of hot spots orbiting JNW strongly naked singularities. As in previous sections, we compare the cases of a Schwarzschild black hole and JNW singularities with $\gamma=0.45$ and $\gamma=0.4$.
\par
We set up a hot spot model, which is an illuminant emitting light isotropically. For the Schwarzschild black hole case with $\gamma=1$, the hot spot is placed at the ISCO, located at $r_{\mathrm{ISCO}}=r_+=6M$. In the case of a JNW strongly naked singularity with $\gamma=0.45$, two marginally stable circular orbits (MSCOs) exist. To capture distinctive observational features, we select the outer MSCO at $r=r_+$. For $\gamma=0.4$, however, no MSCO is present outside the singularity. Following the criterion suggested in \cite{Hamaus:2008yw}, which recommends that hot spots be confined within a compact region near the central object (specifically $r\lesssim 0.3 r_c$), we adopt a representative orbit at $r=6M$ to ensure the hot spot remains within the physically relevant region of the spacetime. We employ the computational framework presented in \cite{Hamaus:2008yw, Rosa:2022toh, Rosa:2023qcv, Chen:2023knf,Wu:2024aa}, with observer positioned at $x^{\mu}_{obs}=(t, r, \theta, \phi)=(t_o, 100M, \theta_o, \pi)$, where the inclination angle $\theta_o$ varies. We generate $500$ snapshots during the orbiting period, and for every snapshot we numerically integrate $1000\times1000$ photon trajectories, each corresponding to a portrayed pixel in the image. At any particular moment $t_k$, we obtain a snapshot image of a lensed hot spot, each pixel of which represents the intensity $I_{klm}$, with the following observable signatures, given in \cite{Rosa:2022toh, Rosa:2023qcv} by,
\begin{itemize}
\item[$\bullet$]
time-integrated images:
\begin{equation}
\langle I\rangle_{lm}=\sum_{k}I_{klm},
\end{equation}
\item[$\bullet$]
total temporal flux:
\begin{equation}
F_k=\sum_l\sum_m\Delta\Omega I_{klm},
\end{equation}
\item[$\bullet$]
temporal magnitude:
\begin{equation}
m_k=-2.5\text{lg}\left(\frac{F_k}{\text{min}(F_k)}\right),
\end{equation}
\item[$\bullet$]
temporal centroid:
\begin{equation}
\vec{c}_k=F^{-1}_k\sum_l\sum_m\Delta\Omega I_{klm}\vec{r}_{lm},
\end{equation}
\end{itemize}
where $\Delta\Omega$ and $\vec{r}_{lm}$ represent, respectively, the pixel solid angle and the position relative to the image centre.
\par
\begin{figure}[htbp]
\centering
\subfigure
{
\begin{minipage}[b]{.3\linewidth}
\centering
\includegraphics[scale=0.3]{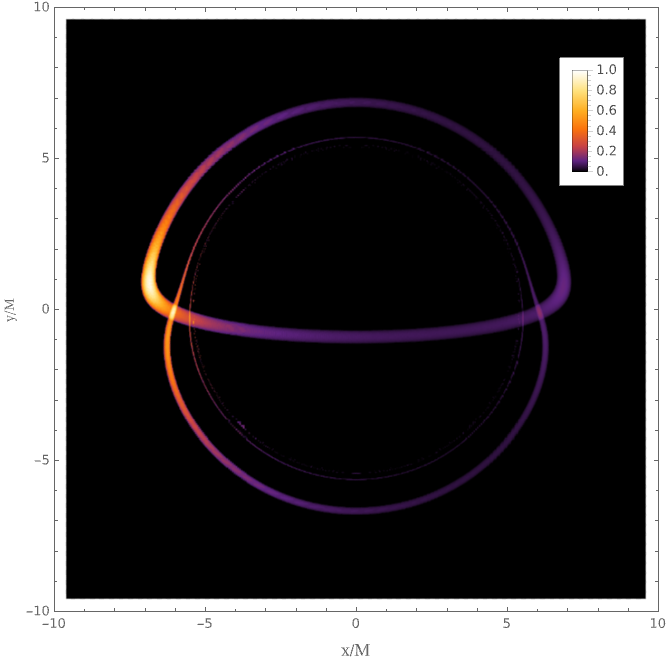}
\end{minipage}
}
\subfigure
{
\begin{minipage}[b]{.3\linewidth}
\centering
\includegraphics[scale=0.3]{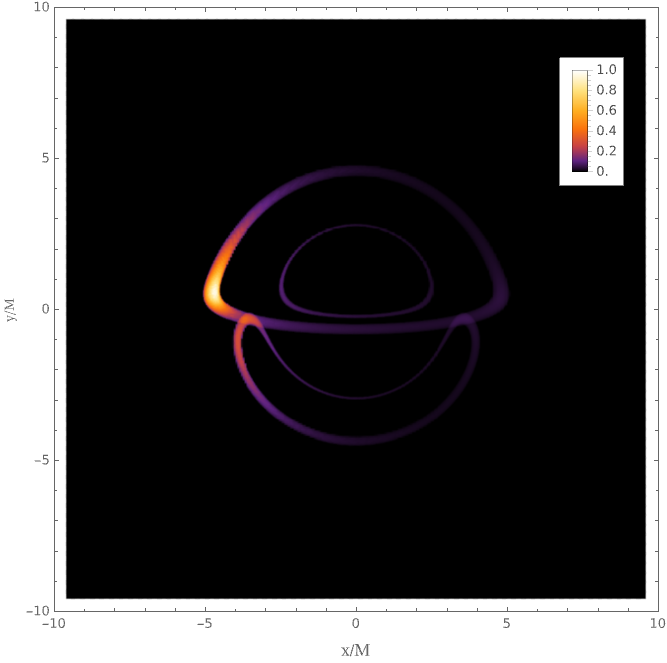}
\end{minipage}
}
\subfigure
{
\begin{minipage}[b]{.3\linewidth}
\centering
\includegraphics[scale=0.3]{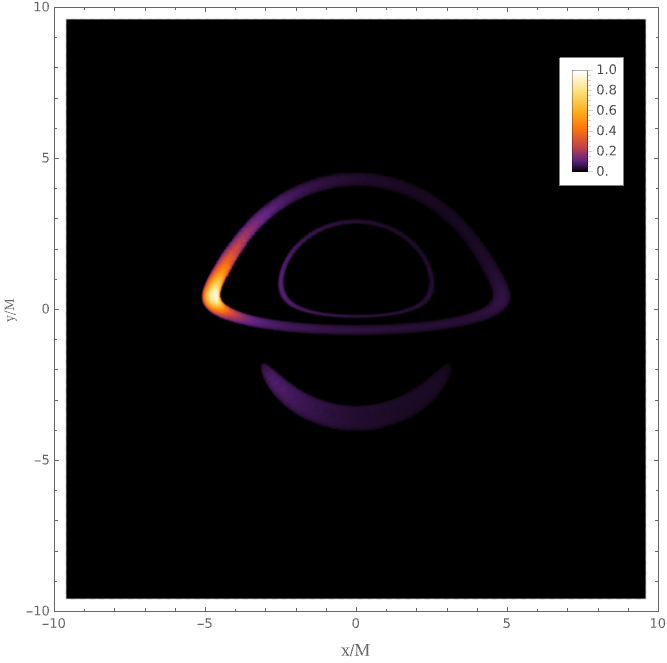}
\end{minipage}
}
\subfigure
{
\begin{minipage}[b]{.3\linewidth}
\centering
\includegraphics[scale=0.3]{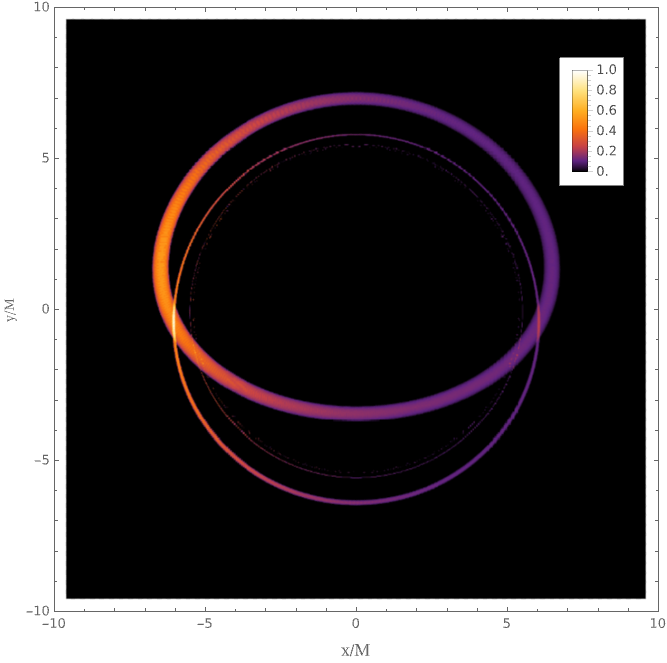}
\end{minipage}
}
\subfigure
{
\begin{minipage}[b]{.3\linewidth}
\centering
\includegraphics[scale=0.3]{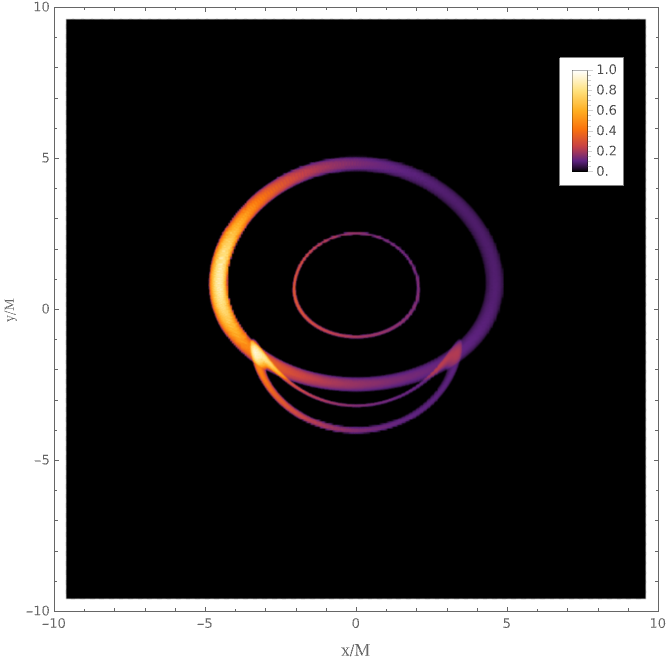}
\end{minipage}
}
\subfigure
{
\begin{minipage}[b]{.3\linewidth}
\centering
\includegraphics[scale=0.3]{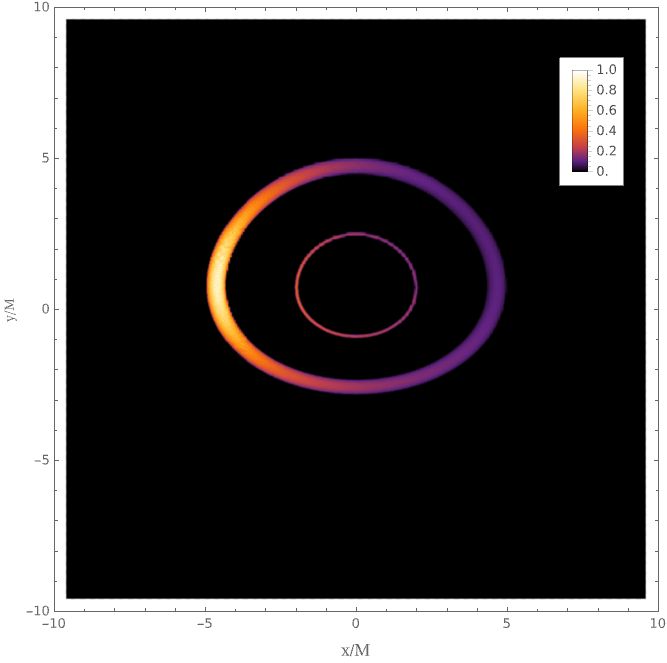}
\end{minipage}
}
\caption{Time integrated images of a full orbit of the hot spot generated from an inclination angle of $\theta_o=80^{\circ}$ (\textbf{Upper Row}) and $\theta_o=50^{\circ}$ (\textbf{Lower Row}), with the magnitude normalized in respect of their maximum value. \textbf{Left Column:} Schwarzschild black hole, with the hot spot positioned at $r=r_{ISCO}$. This column depicts the primary and secondary image outside the photon sphere, produced by the $n=0^{\mathcal{N}}$ and $1^{\mathcal{N}}$ light rays from the hot spot. \textbf{Middle Column:} JNW strongly naked singularity with $\gamma=0.45$, with hot spot situated at $r=r_+$. Both image in this column displays three unique lensed hot spot image tracks, each generated by a distinct process, denoted as $n=0^{\mathcal{N}}$, $1^{\mathcal{N}}$ and $1^{\mathcal{R}}$, and $0^{\mathcal{R}}$ light rays. \textbf{Right Column:} JNW strongly naked singularity with $\gamma=0.4$, with hot spot situated at $r=6M$. Unlike the $\gamma=0.45$ case, the upper row of this column exhibits three lensed images, while only two are depicted in the lower row. The missing image track in the lower-right panel corresponds to that in the upper-right panel, resulting from $n=1^{\mathcal{N}}$ and $1^{\mathcal{R}}$ light rays emitted from the hot spot.}
\label{Figure 6}
\end{figure}

To illustrate the characteristics of the hot spot imaging, their time-integrated images are presented in FIG. \ref{Figure 6}. These images consist of multiple image tracks corresponding to the lensed hot spot observed at different moments along its orbit. To distinguish these tracks, we assign the integer $n$ to denote the number of times a light ray intersects the equatorial plane. Furthermore, since light rays can be reflected in the vicinity of the strongly naked singularity, we introduce subscripts $\mathcal{R}$ and $\mathcal{N}$ to indicate whether the light rays undergo reflection or propagate without reflection, respectively. Notably, the $\mathcal{N}$-type light rays, which are not reflected, correspond to weak gravitational lensing effects in the absence of a photon sphere, as their deflection occurs solely due to the curved spacetime geometry without the confinement of unstable circular photon orbits.
\par
The left column of FIG. \ref{Figure 6} shows the time-integrated images of a hot spot orbiting a Schwarzschild black hole at $r = r_{\mathrm{ISCO}}$. In both panels, the primary and secondary image tracks are visible, along with additional higher-order images that reflect the presence of a photon sphere. The pronounced asymmetry in brightness between the left and right sides of each image arises from the relativistic Doppler effect. In the upper-left panel, corresponding to an inclination angle of $\theta_o = 80^{\circ}$, the semicircular primary image is produced by $n=0^{\mathcal{N}}$ light rays. The secondary image, generated by $n=1^{\mathcal{N}}$ light rays, appears as a smaller, dimmer, and more circular track surrounding the primary one. The lower-left panel, with an inclination angle of $\theta_o = 50^{\circ}$, exhibits a qualitatively similar structure, though the Doppler-induced brightness asymmetry is noticeably reduced at the lower inclination.
\par
The middle column of FIG. \ref{Figure 6} displays the time-integrated images of a hot spot orbiting a JNW strongly naked singularity with $\gamma=0.45$. Unlike the Schwarzschild case, the upper segment of the secondary image is bent down. Instead, another one much fainter image track emerges.  As we analysed in the previous section, owing to the reflection property of the strongly naked singularity, we obtain:
\begin{itemize}
\item[$\bullet$]
The primary image track is formed by $n=0^{\mathcal{N}}$ light rays.
\item[$\bullet$]
The secondary image track is composed of upper and lower segments, each corresponding to $n=1^{\mathcal{N}}$ and $1^{\mathcal{R}}$ light rays, respectively. Both are emitted from the hot spot behind the JNW naked singularity, thus contributing to the flip of the image track generated by the hot spot in front of the singularity.
\item[$\bullet$]
The additional image track generated by $n=0^{\mathcal{R}}$ light rays lies inside the primary image and appears fainter due to the increased apparent angular velocity of the reflected rays near the singularity.
\end{itemize}
As the inclination angle decreases from $\theta_o=80^{\circ}$ to $50^{\circ}$, the brightness asymmetry caused by the Doppler effect becomes less pronounced. Meanwhile, the primary image tracks of $n=0^{\mathcal{N}}$ and $0^{\mathcal{R}}$ appear more circular, and the secondary image tracks of $n=1^{\mathcal{N}}$ and $1^{\mathcal{R}}$ move closer to each other.
\par
In the case of $\gamma=0.4$, shown in the right column of FIG. \ref{Figure 6}, the primary image tracks of $n=0^{\mathcal{N}}$ and $0^{\mathcal{R}}$ remain essentially consistent with those in the $\gamma=0.45$ case. However, the behaviour of the secondary images differs. In the upper-right panel with an inclination angle of $\theta_o=80^{\circ}$, the image tracks produced by $n=1^{\mathcal{N}}$ and $1^{\mathcal{R}}$ overlap, forming a single pattern. In contrast, in the lower-right panel, the secondary image track disappears entirely due to reflection, while the $n=0^{\mathcal{R}}$ track remains visible. This indicates that images of $n^{\mathcal{N}}$ and $n^{\mathcal{R}}$ always emerge in pairs, with $n^{\mathcal{N}}$ consistently located outside $n^{\mathcal{R}}$, provided the reflection occurs.
\par
\begin{figure}[htbp]
\centering
\subfigure
{
\begin{minipage}[b]{.22\linewidth}
\centering
\includegraphics[scale=0.335]{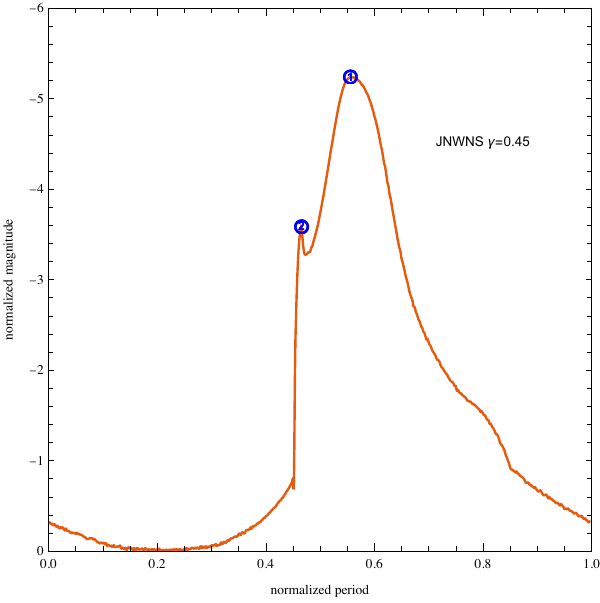}
\end{minipage}
}
\subfigure
{
\begin{minipage}[b]{.22\linewidth}
\centering
\includegraphics[scale=0.35]{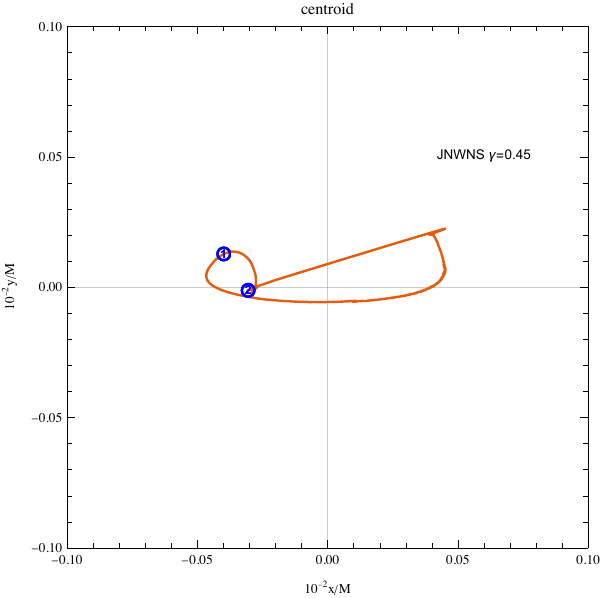}
\end{minipage}
}
\subfigure
{
\begin{minipage}[b]{.22\linewidth}
\centering
\includegraphics[scale=0.2]{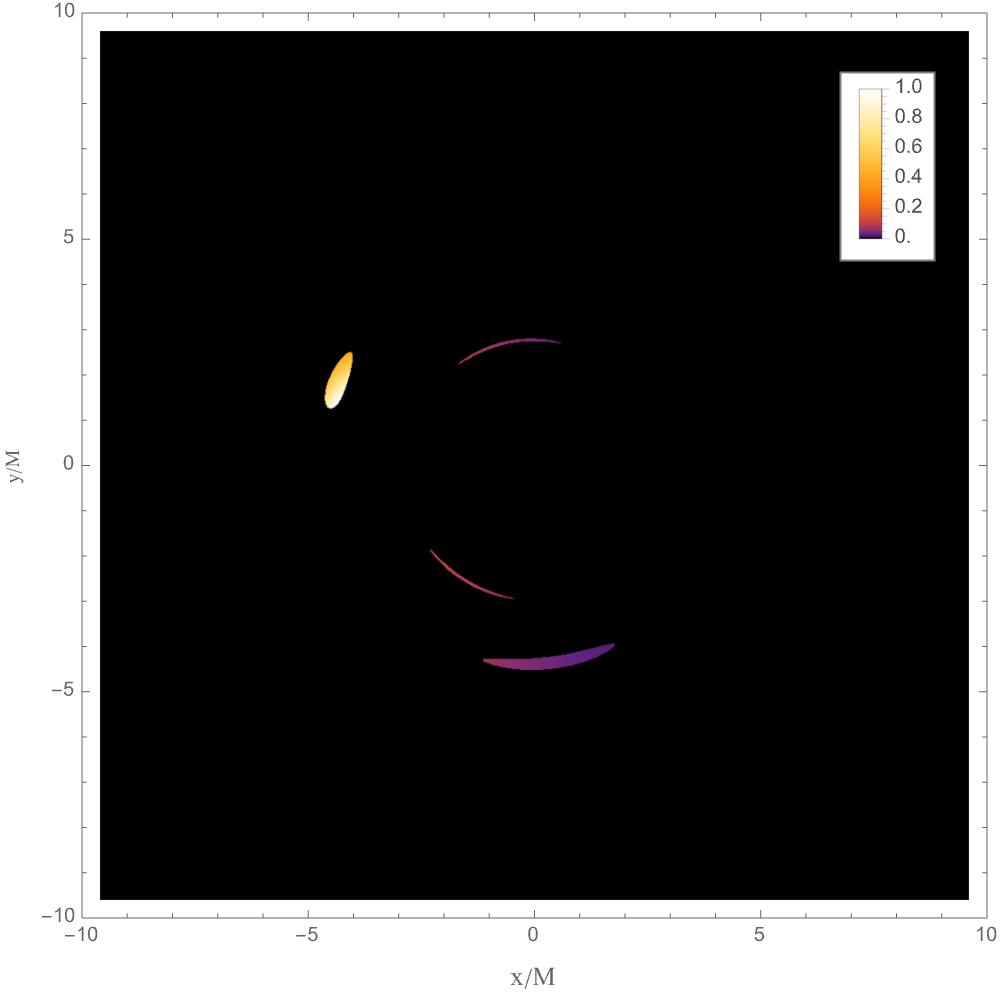}
\end{minipage}
}
\subfigure
{
\begin{minipage}[b]{.22\linewidth}
\centering
\includegraphics[scale=0.2]{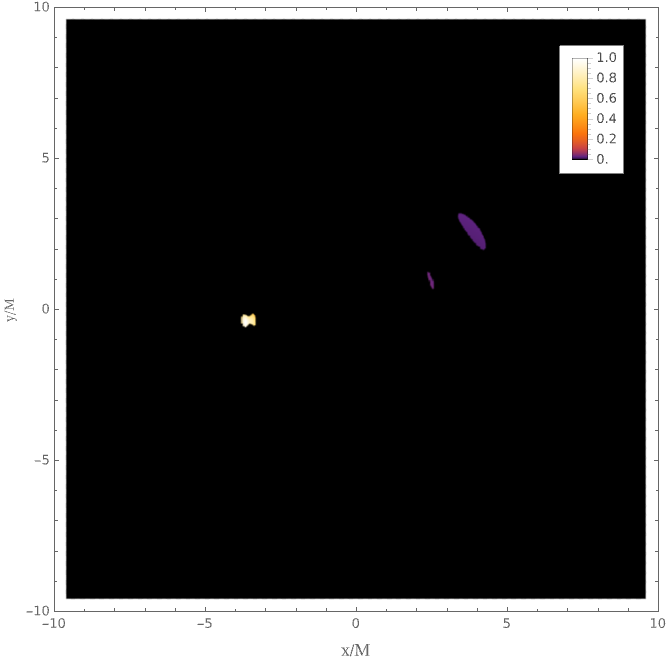}
\end{minipage}
}
\subfigure
{
\begin{minipage}[b]{.22\linewidth}
\centering
\includegraphics[scale=0.335]{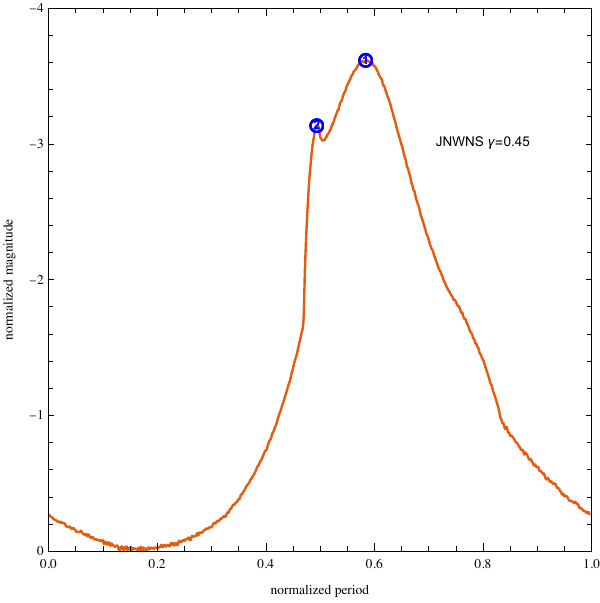}
\end{minipage}
}
\subfigure
{
\begin{minipage}[b]{.22\linewidth}
\centering
\includegraphics[scale=0.35]{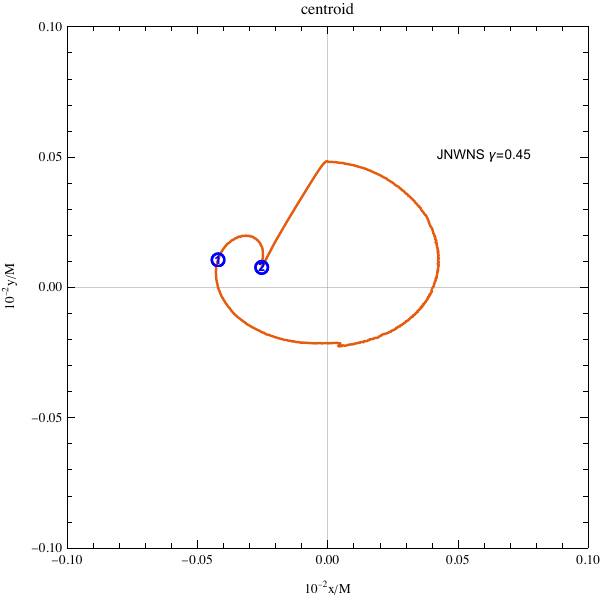}
\end{minipage}
}
\subfigure
{
\begin{minipage}[b]{.22\linewidth}
\centering
\includegraphics[scale=0.2]{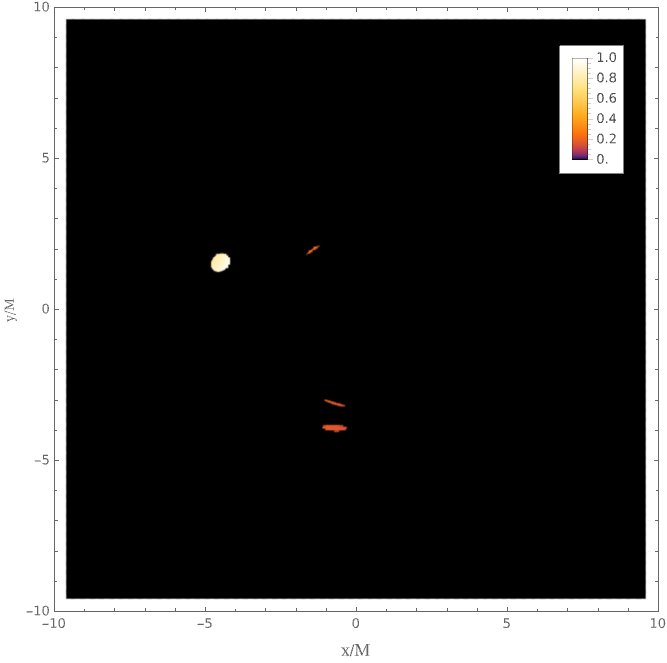}
\end{minipage}
}
\subfigure
{
\begin{minipage}[b]{.22\linewidth}
\centering
\includegraphics[scale=0.2]{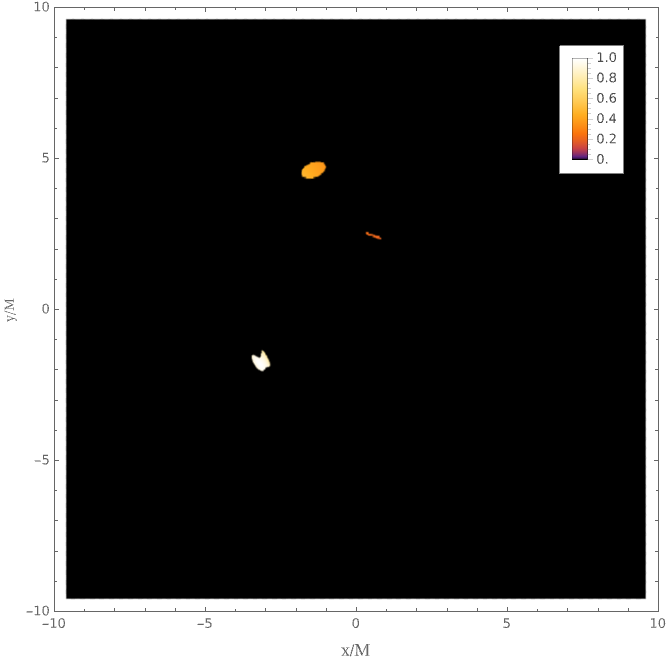}
\end{minipage}
}
\caption{Temporal magnitude $m_k$ (\textbf{First Column}) and centroids $\vec{c}_k$ (\textbf{Second Column}) as a function of normalized period in the case of $\gamma=0.45$ with the inclination angle $\theta_o=80^{\circ}$ (\textbf{Upper Row}) and $50^{\circ}$ (\textbf{Lower Row}). Snapshot images at the peak of the temporal magnitude are attached as well (\textbf{Third Column} and \textbf{Fourth Column}). In both scenarios, two peaks are observed, with the highest and secondary peaks labelled as $\textcircled{\scriptsize{1}}$ and $\textcircled{\scriptsize{2}}$, accompanied by the corresponding centroids.}
\label{Figure 7}
\end{figure}

To further investigate the dynamic characteristics of the lensed hot spot images around JNW strongly naked singularities, we examine the temporal magnitude $m_k$ and centroid trajectories $\vec{c}_k$. The analysis is performed for two representative cases, $\gamma=0.45$ and $0.4$, each at inclination angles of $\theta_o=80^{\circ}$ and $50^{\circ}$. Significant peaks in the temporal magnitude curves are marked with numerical labels, with corresponding image snapshots provided for clarity. FIG. \ref{Figure 7} presents the results for $\gamma=0.45$, where the hot spot orbits at $r=r_+$. In both inclination scenarios, the primary peak in $m_k$ arises when the hot spot moves to the far left of its orbit, with the image formed by $n=0^{\mathcal{N}}$ light rays. This brightness enhancement is attributed to the relativistic Doppler effect, as shown in the third column of FIG. \ref{Figure 7}. When the hot spot reaches the rightmost position relative to the observer, the temporal magnitude exhibits a secondary peak. This is predominantly contributed by the secondary images produced by $n=1^{\mathcal{N}}$ and $n=1^{\mathcal{R}}$ light rays, as illustrated in the fourth column. The appearance of these distinct peaks reflects the combined effects of gravitational lensing and Doppler beaming in the absence of a photon sphere, modulated by both the orbital phase and the inclination angle.
\par
\begin{figure}[htbp]
\centering
\subfigure
{
\begin{minipage}[b]{.3\linewidth}
\centering
\includegraphics[scale=0.435]{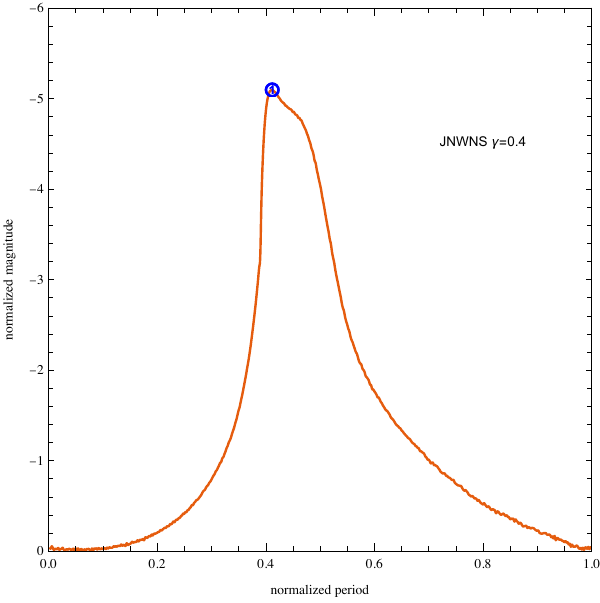}
\end{minipage}
}
\subfigure
{
\begin{minipage}[b]{.3\linewidth}
\centering
\includegraphics[scale=0.45]{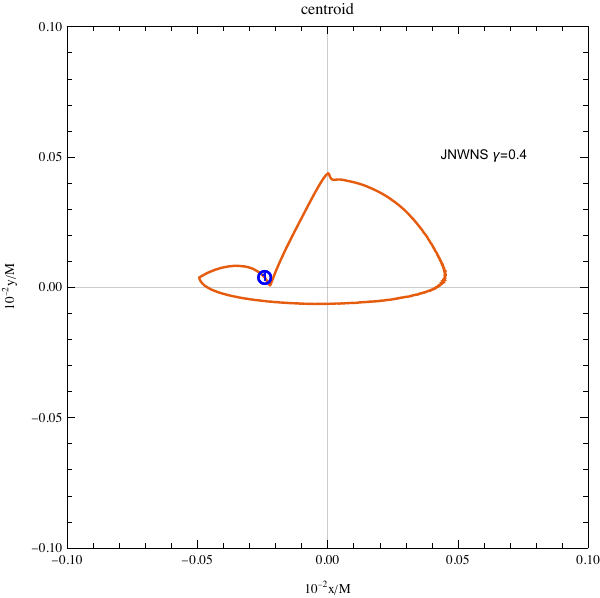}
\end{minipage}
}
\subfigure
{
\begin{minipage}[b]{.3\linewidth}
\centering
\includegraphics[scale=0.25]{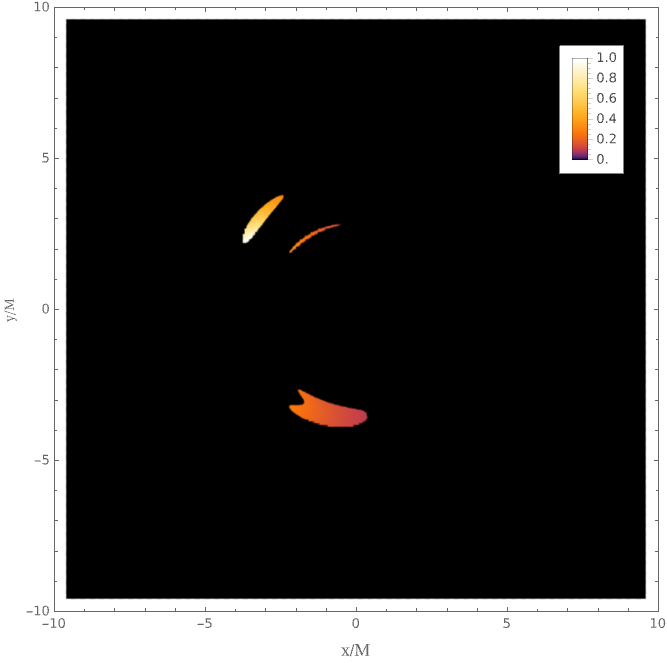}
\end{minipage}
}
\subfigure
{
\begin{minipage}[b]{.3\linewidth}
\centering
\includegraphics[scale=0.435]{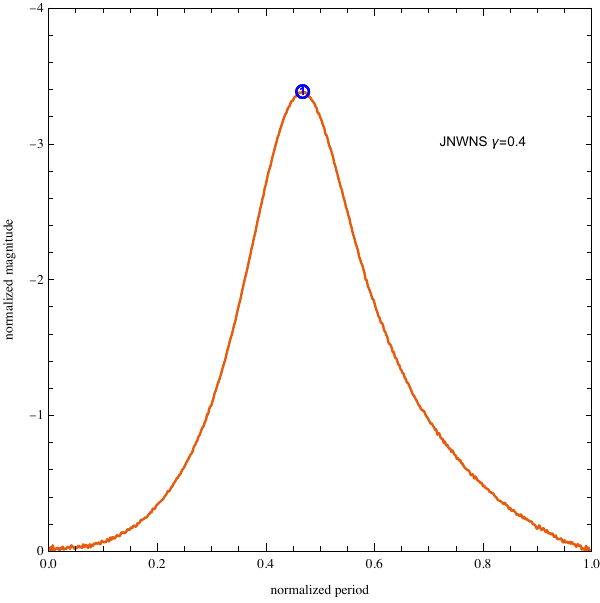}
\end{minipage}
}
\subfigure
{
\begin{minipage}[b]{.3\linewidth}
\centering
\includegraphics[scale=0.45]{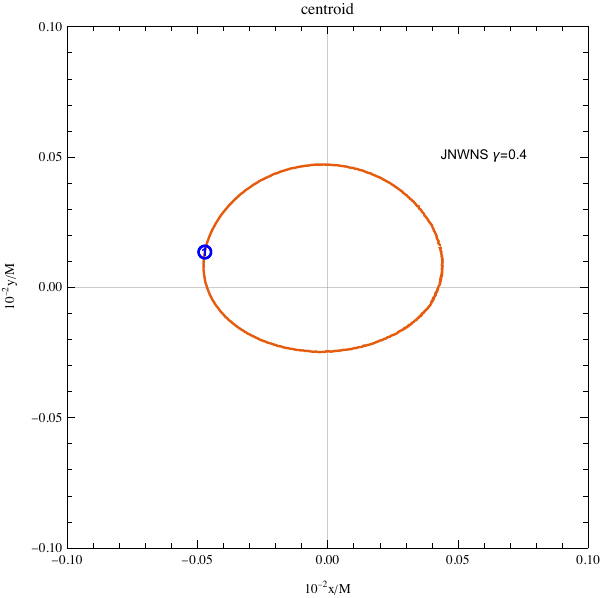}
\end{minipage}
}
\subfigure
{
\begin{minipage}[b]{.3\linewidth}
\centering
\includegraphics[scale=0.25]{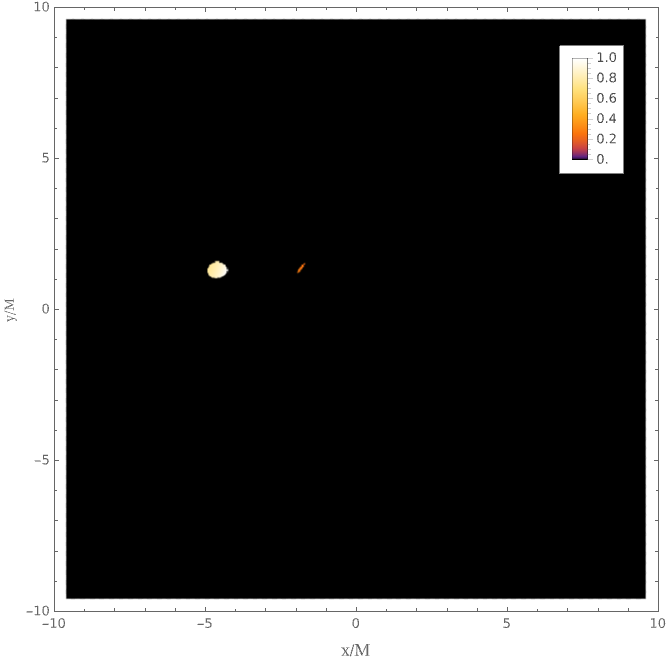}
\end{minipage}
}
\caption{Temporal magnitude $m_k$ (\textbf{Left Column}) and centroids $\vec{c}_k$ (\textbf{Middle Column}) as a function of normalized period in the case of $\gamma=0.4$ with the inclination angle $\theta_o=80^{\circ}$ (\textbf{Upper Row}) and $50^{\circ}$ (\textbf{Lower Row}). Snapshot images at the peak of the temporal magnitude are attached as well (\textbf{Right Column}). In both scenarios, only one peak of temporal magnitude is observed, which is labelled as $\textcircled{\scriptsize{1}}$, along with the centroid of the flux at the peak.}
\label{Figure 8}
\end{figure}
\par
FIG. \ref{Figure 8} displays the temporal magnitude $m_k$ and centroid trajectories $\vec{c}_k$ for hot spots orbiting the JNW naked singularity with $\gamma=0.4$ at $r=6M$, for inclination angles of $\theta_o=80^{\circ}$ and $50^{\circ}$. In contrast to the $\gamma=0.45$ case, only a single prominent peak appears in the temporal magnitude curves for both inclination angles, corresponding to the primary image produced by $n=0^{\mathcal{N}}$ light rays, as indicated in the right column. As revealed by the centroid tracks and image snapshots, for $\theta_o=80^{\circ}$, the secondary images formed by $n=1^{\mathcal{N}}$ and $n=1^{\mathcal{R}}$ nearly overlap, effectively merging into a single track. This overlap suppresses the formation of a distinct secondary peak in $m_k$. At a lower inclination of $\theta_o=50^{\circ}$, the reflection effect further leads to the complete disappearance of the secondary image, leaving only the primary track and its associated peak. These results consistently reflect the influence of both the reflection-induced image inversion and inclination-dependent lensing behavior in the absence of a photon sphere.

\section{Conclusions}

\label{sec:CONCLUSIONS}

In this paper, we investigated the observational signatures of the spacetime around the JNW strongly naked singularity, comprising the celestial sphere, accretion disk, and hot spots orbiting the singularity. JNW strongly naked singularity indicates that it is uncovered by any horizon or photon sphere, which contributes to a significant distinction in observable properties. In each part, we first establish the framework of the observations, based on the stable circular orbits in scenarios where $\gamma$ varies. Subsequently, methods of ray-tracing and numerical simulation are employed to generate lensing images. To make the observed properties more intuitive, we also present the case of a Schwarzschild black hole together with a JNW strongly naked singularity.
\par
Our findings demonstrate that the JNW strongly naked singularity exhibits a reflective characteristic, causing light rays traveling in its vicinity to undergo reflection, which in turn generates the inversion of the secondary image. This reflective behavior also produces additional images. Regardless of the structure of the accretion disk, the resulting images consistently include a fainter component and a series of ring-shaped features within it. Moreover, through the analysis of lensed hot spot images, we identified not only the inversion of the secondary image track but also the appearance of an additional reflected track. Notably, image tracks corresponding to $n^{\mathcal{N}}$ and $n^{\mathcal{R}}$ always appear in pairs, arranged in a thin bar-shaped structure, with the $n^{\mathcal{N}}$ track consistently located outside the $n^{\mathcal{R}}$ track unless the inversion occurs. Importantly, excessive inversions suppress the formation of the secondary peak in the temporal magnitude.
\par
The aforementioned analysis offers a valuable asset in discerning JNW strongly naked singularities. Furthermore, it catalyses research aimed at probing the essence of naked singularities and exploring alternatives to black holes. The emergence of the next-generation Event Horizon Telescope (EHT) heralds an era of amplified observational capabilities, allowing for the scrutiny of finer details. Thus, further investigations with this work will contribute to substantiating the existence of a naked singularity.

\begin{acknowledgments}

We are grateful to Guangzhou Guo and Yiqian Chen for useful discussions and valuable comments. This work is supported in part by NSFC (Grant No. 12275183, 12275184 and 11875196). 

\end{acknowledgments}

\bibliographystyle{unsrturl}
\bibliography{ref}

\end{document}